\documentclass[conference]{IEEEtran}

\usepackage{cite}
\usepackage{amsmath,amssymb,amsfonts}
\usepackage{algorithmic}
\usepackage{graphicx}
\usepackage{textcomp}
\usepackage{xcolor}
\usepackage{multicol}
\usepackage{flushend}

\def\BibTeX{{\rm B\kern-.05em{\sc i\kern-.025em b}\kern-.08em
    T\kern-.1667em\lower.7ex\hbox{E}\kern-.125emX}}

\begin{document}

\IEEEoverridecommandlockouts
\IEEEpubid{\begin{minipage}[t]{\textwidth}\ \\[8pt]
        \normalsize{The work presented here was originally published at the 7th Embedded Security in Cars Conference (ESCAR USA 2019). Copyright 2021 Intel Corporation. Distribution permission granted to arXiv.org under the terms of the following license: https://arxiv.org/licenses/nonexclusive-distrib/1.0/license.html.}
\end{minipage}}

\title{Two-Point Voltage Fingerprinting: Increasing Detectability of ECU Masquerading Attacks
}

\author{
\IEEEauthorblockN{Shabbir Ahmed}
\IEEEauthorblockA{
\textit{Intel Corporation}\\
Hillsboro, Oregon USA\\
shabbir.ahmed@intel.com} \\
\and
\IEEEauthorblockN{Marcio Juliato}
\IEEEauthorblockA{
 \textit{Intel Corporation}\\
Hillsboro, Oregon USA \\
marcio.juliato@intel.com}
\and
\IEEEauthorblockN{Christopher N. Gutierrez}
\IEEEauthorblockA{
\textit{Intel Corporation}\\
Hillsboro, Oregon USA \\
christopher.n.gutierrez@intel.com}
\and
\IEEEauthorblockN{Manoj Sastry}
\IEEEauthorblockA{
\textit{Intel Corporation}\\
Hillsboro, Oregon USA \\
manoj.r.sastry@intel.com}
}

\maketitle

\begin{abstract}
Automotive systems continuously increase their dependency on Electronic Control Units (ECUs) and become more interconnected to improve safety, comfort and Advanced Driving Assistance Systems (ADAS) functions to passengers and drivers. As a consequence of that trend, there is an expanding attack surface which may potentially expose vehicle's critical functions to cyberattacks. It is possible for an adversary to reach the underlying Control Area Network (CAN) through a compromised node or external-facing network interface, and launch masquerading attacks that can compromise road and passenger safety. Due to lack of native authentication in the CAN protocol, an approach to detect masquerading attacks is to use ECU voltage fingerprinting schemes to verify that the messages are sent by authentic ECUs. Though effective against simple masquerading attacks, prior work is unable to detect attackers such as hardware Trojans, which can mimic ECU voltages in addition to spoofing messages. We introduce a novel \textit{Two-point ECU Fingerprinting} scheme and demonstrate efficacy in a controlled lab setting and on a moving vehicle. Our results show that our proposed two-point fingerprinting scheme is capable of an overall F1-score over 99.4\%. The proposed approach raises the bar for attackers trying to compromise automotive security both remotely and physically, therefore improving security and safety of autonomous vehicles.
\end{abstract}


\section{Introduction}

Over the course of the last few decades, automotive systems have evolved from purely mechanical to Cyber-Physical Systems (CPS). An invariant in the overhaul of automotive design has been road and passenger safety. Electronic Control Units (ECUs) are fundamental building blocks to deliver more complex safety features such as Advanced Driving Assistance Systems (ADAS), as well as to deliver safer autonomous vehicles. The automotive industry has also been transforming vehicles into interconnected nodes, which are now capable of reaching the outside world through cellular and Wi-Fi networks, and will soon take advantage of Vehicle-to-Everything (V2X) capabilities. Consequently, on-board systems that used to be fully isolated can now be easily reachable from the Internet. The resulting scenario is an expanded attack surface, leading to an increased exposure to cyber-threats. Hence, security becomes yet another requirement to be properly addressed, otherwise electronic devices originally conceived to increase safety can now undermine their main function.

Even with all the advances of on-board electronics along with the constantly evolving security scene, the vast majority of modern vehicles still rely on the CAN bus~\cite{bosch_canspec} as the de-facto In-Vehicle Network (IVN). The CAN bus is a broadcast-based protocol that was devised in the end of the 80's, which uses Message Identifiers (MIDs) to resolve bus arbitration and to identify the message being transmitted. Though it meets high reliability requirements, it inherently lacks fundamental security features such as data origin authentication. Any node connected to the bus can inject frames with any MID, making it trivial to masquerade as authentic ECUs. Receiving nodes cannot distinguish whether a frame came from an authentic ECU of from an attacker. For instance, an attacker who compromised an Infotainment Unit (IVI) over the Internet can cause serious safety consequences by injecting safety-critical commands (e.g. steering, acceleration, and braking), while remaining undetected. Currently, this is one of the weakest links in automotive security.

During the last several years, security researchers have raised awareness of these automotive security issues~\cite{KCRPKCMKASS10,CMKASSKCRK11,HKD11,MV13,MV14}, and more recently, attacks have been demonstrated against real vehicles~\cite{MV15,Greenberg15,KL15,KSL16,KSL17,Kovelman17}. For instance, in the infamous Jeep hack incident~\cite{MV15,Greenberg15,KL15}, researchers remotely tampered with air conditioning, windshield wipers, radio, and ultimately killed the engine in the middle of a highway. The entry point was a vulnerable cellular connection in the vehicle, which allowed the attackers to reprogram the IVI firmware to ultimately reach the CAN bus and perform malicious actions. Tesla vehicles were also targets of attacks~\cite{KSL16,KSL17}. Researchers exploited vulnerabilities in the vehicle's Wi-Fi system to demonstrate attacks against exterior lights, seat configuration, doors, and ultimately applied the brakes in a moving vehicle. Even though Tesla responded with security updates~\cite{Greenberg16}, new vulnerabilities were exploited in~\cite{KSL17}. Again, the researchers were able to control doors, brakes, and to present a coordinated attack that they named "the unauthorized Xmas show".

The commonality among all the aforementioned attacks is masquerading of ECUs connected to the CAN bus. It is impractical to apply authentication mechanisms directly to the CAN bus due to legacy latency and bandwidth constraints, as well as key management issues. A number of methods have been proposed in the literature~\cite{HKD11,MA11,SKK16,MBCSP17} as a countermeasure against masquerading attacks. The ones in~\cite{CS16,MG14,CS17,AHTM17,CJJPL18} are based on fingerprinting the signal characteristics of each ECU. We refer to those as methods as {\it Single-Point Fingerprinting}. Those schemes can serve as a building block for Intrusion Detection Systems (IDSs). Though they can offer some level of protection, single-point IDSs schemes are unable to defend against more advanced attacks. For example, hardware Trojans in an ECU's CAN interface or malicious devices connected to the On-board Diagnostics (OBD-II) port could inject malicious messages whose voltage profiles match those of authentic ECUs, therefore bypassing the aforementioned protection mechanisms. 

Hardware Trojans have been extensively studied ~\cite{ChipHack, SAM13, LI2016426, BOZDAL201847}. Recently, Amazon Web Services\texttrademark\ found a hardware Trojan deployed as a tiny microchip on a server motherboard which was capable of creating a stealth doorway into the network that was not part of the original design~\cite{ChipHack}. In the automotive domain, researchers have shown how a hardware Trojan in an CAN interface can lead to denial of service without an adversary having physical access to the CAN bus~\cite{BOZDAL201847}. This type of a pervasive hardware-based attacks are not only hard to detect but can potentially have devastating consequences. Since the automotive ecosystem relies on various ECU providers, compromised CAN transceivers (such as proposed in ~\cite{BOZDAL201847}) can have a widespread impact on the security of in-vehicle networks. By relying on hardware Trojans, attackers could have a solid foothold to launch more advanced attacks to circumvent existing techniques. For instance, compromised transceivers could launch masquerading attacks by spoofing voltage of authentic ECUs (as described in~\cite{CS17,CJJPL18}) and thus bypass single-point fingerprint mechanisms.

In addition, OBD-II ports are mandatory for all vehicles in the US~\cite{ISO_15765_4} and it are typically accessible under the dashboard. Hence, it can provide access to the CAN bus to benign (e.g. manufacturers, service maintenance, etc.) and malicious actors. Recently, an attack against an OBD-II Bluetooth dongle has been published~\cite{Kovelman17}. In this attack, the Bluetooth PIN was brute-forced which consequently allowed the attacker to inject messages directly into the CAN bus to turn off the vehicle's engine. Physical access to the vehicle subsystems is a reality in valet parking, rental cars, ride-sharing vehicles, self-driving taxis, and fully autonomous vehicles. Once an attack device is plugged to OBD-II, it can serve as a foothold for remote actors to tamper with critical controls (e.g. steering, braking). Resilience against those kinds of attacks must be provided to achieve proper vehicular security, and consequently passenger and road safety.

The main contributions of this paper are as follows:
\begin{itemize}
	\item An in-depth analysis of single-point fingerprinting schemes is performed, and a practical attack against those schemes is presented. We demonstrate that with inexpensive hardware, an attacker can equip current masquerading attack modules to bypass voltage fingerprinting schemes.
	\item A novel Two-Point voltage fingerprinting is introduced, which utilizes the strengths of single-point fingerprinting, while being resistant to more advanced attackers that can inject malicious signals into the CAN network.
	\item Demonstrated the feasibility of the proposed approach with an F1-score of more than 99.4\% on an ECU testbed and on an actual vehicle.
\end{itemize}

The remainder of this paper is organized as follows: Section~\ref{sec:background} reviews fundamental concepts on the CAN bus. In Section~\ref{sec:ttm} we describe the threat model, assumptions, and details regarding an attack to bypass single point ECU fingerprinting schemes such as the ones found in ~\cite{CS16,MG14,CS17,AHTM17,CJJPL18}. Next, in Section~\ref{sec:2p_vf} we introduce a novel Two-Point ECU fingerprinting approach that is resistant to MID-Voltage masquerading attacks, and in Section~\ref{sec:exp_results} we demonstrate its feasibility and efficacy. We provide a comprehensive discussion on the advantages of our approach and compare related work in Section~\ref{sec:comp_related_work}. Section~\ref{sec:conclusions} presents our conclusions.

\section{CAN Bus Background}
\label{sec:background}

The physical CAN bus is comprised of two wires that carry signals CANH and CANL~\cite{bosch_canspec}. The bus interface is usually made of a CAN controller~\cite{microchip_mcp2515} and a CAN transceiver~\cite{microchip_mcp2551}. The former is responsible for the execution of the CAN protocol, while the latter performs the actual interface with the physical layer of the CAN bus. The differential voltages between CANH and CANL will convey whether a logical 0 or 1 is being transmitted. A logical 0 is represented as a dominant signal, where CANH would output 3.5V and CANL 1.5V. A logical 1 is represented as a recessive signal, whose nominal voltages are approximately 2.5V for both CANH and CANL.

In practice, recessive levels can vary between 2 and 3V, granted a minimum and maximum differential tolerance of \mbox{-500mV} and 50mV, respectively. Dominant levels on CANH can vary from 2.75 and 4.5V, whereas CANL can vary between 0.5 to 2.25V. The minimum and maximum differential dominant output voltages are respectively 1.5 and 3V. Actual output voltages from each ECU in the CAN bus will vary within the aforementioned limits and is a feature used in voltage fingerprinting. The dominant and recessive terminology represents how the bus levels settle when multiple nodes are trying to drive it at the same time. In other words, dominant signals will override the bus level over recessive signals.

\begin{figure}[!htb]
\begin{center}
\includegraphics[width=0.48\textwidth]{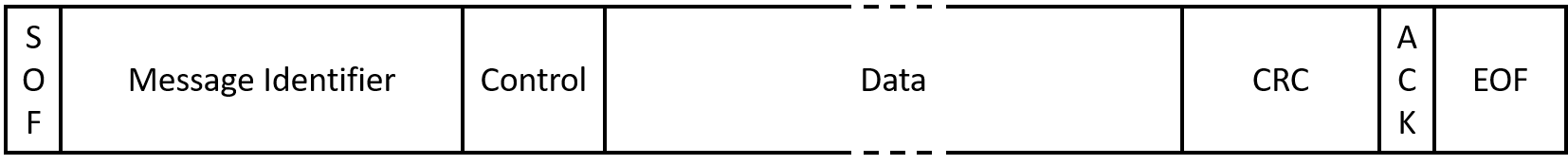}
\caption{Standard CAN Frame and Fields}
\label{fig:can_frame}
\end{center}
\end{figure}

The CAN standard~\cite{bosch_canspec} specifies a series of frames, namely: data (standard and extended), remote, overload, error frames. The most commonly used type of frame is the standard data frame, which carries the payload from transmitters to receivers. As depicted in Figure~\ref{fig:can_frame}, the data frame main fields are: arbitration (11 bits), control (6 bits), data (0 to 64 bits), CRC (15 bits) and ACK (acknowledge from other nodes in the bus). There are some additional fields, whose are listed in a more detailed description of the CAN frame provided at~\cite{bosch_canspec,microchip_mcp2515}.

Bit stuffing is one of the fault detection mechanisms in the CAN protocol~\cite{bosch_canspec}, which specifies that a bit of opposite polarity must be placed after a sequence of five consecutive bits of the same polarity. This is transparent to the transmitter/receiver since it is automatically performed by the CAN protocol chip.

The arbitration field is interchangeably called Message ID (MID) field, due to its dual function in the CAN bus. The protocol does not enforce any restriction on the MIDs a given node can transmit. In practice, automotive designers assign a set of MIDs for each ECU connected to the bus; under normal conditions a specific ECU will only transmit the MIDs it has been assigned.

\section{Threat and Attack Models}
\label{sec:ttm}

In order to clearly outline the attacker profile addressed in this work, the following threat model is considered:
\begin{itemize}
    \item An ECU on the bus can send messages with any MID at the discretion of the attacker, therefore being able to masquerade any other node on the bus.
    \item Attackers can inject packages with modulated signals, hence being able to spoof voltage of authentic ECUs, e.g. through hardware Trojans, devices connected to OBD-II.
    \item The attacker does not disconnect ECUs or suspend messages.
\end{itemize}

As described in the previous section, the CAN protocol does not have message authentication built into it, which allows a compromised ECU to masquerade as a different. In this paper, two main attacks are considered: 1) (Traditional) MID Masquerading, 2) Masquerading with both MID and Voltage (denoted as \mbox{MID-Voltage Masquerading}). 

    
    
\subsection{MID Masquerading}

Traditional MID masquerading are very well documented in the literature, and is the most prevalent type of attack against automotive systems~\cite{MV15,Greenberg15,KL15,KSL16, KSL17,Greenberg16}. In this attack, the attacker gains control of an existing ECU or connects an external node to the CAN bus. Next, it sends out MIDs belonging to other nodes of the network, therefore impersonating them. In that attack model, the masquerading node do not change the output voltage levels of its transceiver. For example, if the compromised ECU\textsubscript{2} of Figure~\ref{fig:attack_types} sends out MIDs of other ECUs (illustrated by MID\textsubscript{n}) its output voltage $V_2$ would still be the characteristic one for ECU\textsubscript{2}'s.

The fundamental observation in voltage fingerprinting schemes is that each ECU outputs a characteristic signal when transmitting messages to the bus 
(refer to Section~\ref{sec:background} for more details). 
The voltage variance, manifested in the order of millivolts, are due to process variations in the transistors of the CAN transceivers within each ECU. In practice, there is enough statistical dispersion in the observed signals to regard them as unique for each ECU. Based on that, it is possible to build a mapping between the observed characteristics and their associated set of MIDs, and later utilize new observations to classify them as one of the known ECUs, and prevent MID Masquerading.

\subsection{MID-Voltage Masquerading Attacks}

This section revisits voltage impersonation attacks, which have already been introduced in literature~\cite{CS17,CJJPL18}. Traditional voltage fingerprinting mechanisms are capable of detecting MID masquerading attacks. However, they are ineffective against more advanced attacks due to the fact that there is no satisfactory freshness and randomness in the signals coming out of the ECUs. As a result, a malicious node can mimic the voltage profile that has been observed in the bus to mislead single-point voltage fingerprinting schemes. In other words, the same way that the stability of the ECUs' voltage profiles enable voltage profiling, they also benefit the attackers. From the attacker's stand point, all messages on the bus are available in clear text, which can later be used in a masquerade attack. In practice, the search space for a matching voltage profile is so small that the attacker does not even have to observe the bus to accumulate knowledge about MIDs and ECU voltages (as described in detail in Section~\ref{subsec:exhaustive_search}).

\begin{figure}[!htb]
\begin{center}
\includegraphics[width=0.5\textwidth]{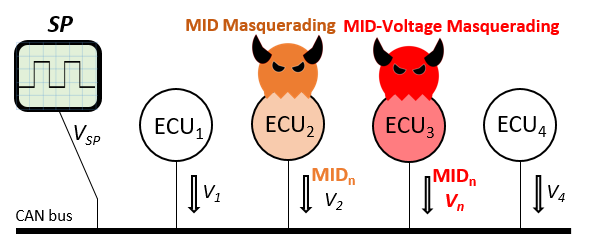}
\caption{MID and MID-Voltage Masquerading Attacks}
\label{fig:attack_types}
\end{center}
\end{figure}

MID-Voltage attacks consider that the malicious node is not only capable of injecting any MIDs, but also voltages pertaining to any other node of the bus (as represented by $V_n$ in Figure~\ref{fig:attack_types}). For example, the attacker can inject malicious messages (MID and Voltage) to masquerade as the Adaptive Cruise Control (ACC) ECU to cause the vehicle to accelerate and/or brake. It is worth noting that this kind of attack would be undetected by existing fingerprinting schemes.
    
With self-driving cars moving towards services (e.g. driverless taxis, ridesharing, etc.), physical access to electronic components of third-party cars is becoming increasingly easier. For instance, physical access to the CAN bus through the OBD-II port typically located near the dash of the car can serve as easy access point to connect malicious devices. Aftermarket devices that add functions like remote start, diagnostic tools, insurance dongles, ADAS functions are also being adopted widely. Some vulnerabilities have already been analyzed in such devices in recent literature~\cite{Kovelman17}.

\subsection{Attack Module for MID-Voltage Masquerading}
\label{sec:attack_module}

\begin{figure}[!htb]
\begin{center}
\includegraphics[width=0.50\textwidth]{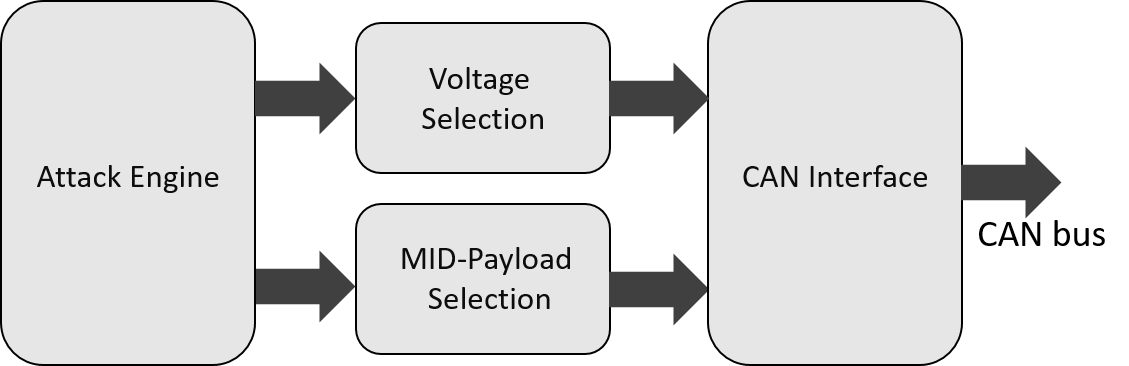}
\caption{Attack Module Block Diagram}
\label{fig:attack_module_block_diagram}
\end{center}
\end{figure}

In order to demonstrate and characterize our approach, we devised an attack module to carry out both MID and \mbox{MID-Voltage} masquerading. The \textit{Attack Engine} executes the masquerading attacks by selecting MIDs, attack payloads, and output voltages as depicted in Figure~\ref{fig:attack_module_block_diagram}.

\subsubsection{Probability of Successful MID-Voltage Masquerading}
\label{subsec:exhaustive_search}

Due to the limited range of operational voltage levels of the CAN bus as described in Section~\ref{sec:background}, the search space for the attacker to guess the voltage level is also low. The search space is bounded by the total number of attack voltage levels $V$ that are possible.

The minimum and maximum voltages for CANH dominant levels are respectively 2.75V and 4.5V, whereas for CANL they are respectively 0.5V and 2.25V. This represents a maximum range $N$ of 1.75V. Assuming a high resolution of approx. 5mV resolution, the number of voltage levels to explore the entire dominant signal tolerance is $L = \frac{N}{5mV}$, therefore $L= 350$.

Given the results above, the attacker has 1 in 350 chance to hit the right voltage level even if it is selected at random. This means it is completely feasible for an attacker to not only inject masqueraded message onto the CAN Bus, but also spoof a voltage profile to overcome existing single-point voltage fingerprinting techniques. Further, the differential cost of implementing such an attack is inexpensive. The high probability of successful attacks and the low cost of implementation highlight the gaps in security of single-point voltage fingerprinting schemes.

A number of related work is presented in Section~\ref{sec:comp_related_work} which rely on utilizing the voltage characteristics of the CAN frame to fingerprint the transmitter ECU. A car in an operational environment is prone to noise on the communication channels and as such, Single Point fingerprinting mechanisms are highly susceptible to this type of noise. In this paper, we introduce a novel Two-Point voltage fingerprinting method that address the aforementioned security concerns.

\section{Two-Point Voltage Fingerprinting}
\label{sec:2p_vf}

To provide resistance against MID-Voltage masquerading attacks, which circumvent detection in prior fingerprinting schemes, we propose a new approach that utilizes the physical characteristics of the bus to uniquely identify each transmitting ECU. We achieve this by observing the intrinsic effects of the bus like resistance, capacitance, and inductance on the transmitted signal. In our approach, we take into account the voltages measured at two points on the bus to determine the location of the transmitting ECU. As a result, our Two-point Voltage Fingerprinting scheme is capable of mapping the unique location of the ECU with the transmitted messages, and therefore detect \mbox{MID-Voltage} masquerading attacks.

\subsection{Trust Model}
In this paper we consider both MID and \mbox{MID-Voltage} masquerade attacks. In comparison to related work reported in Section~\ref{sec:comp_related_work}, the threat model represents a more capable adversary (e.g. hardware Trojan), who may execute \mbox{MID-Voltage} masquerading attacks.
The goal of the proposed work is to detect {\it both} MID and MID-Voltage masquerading attacks, which are not considered in prior work.

Our trust model considers that:
\begin{itemize}
    \item As per the CAN protocol requirements~\cite{bosch_canspec}, properly configured (authentic) ECUs have their own set of MIDs that do not overlap with the other ECUs.
    \item Training of machine learning models is performed in an uncompromised environment (i.e. no adversaries present, and no misbehaving ECUs).
    \item Environmental changes impacts the entire physical channel uniformly.
    \item The voltage fingerprinting mechanism is inaccessible to the adversary and remains in normal operational during attacks.
\end{itemize}

\subsection{Fundamental Principle}

\begin{figure}[!htb]
\begin{center}
\includegraphics[width=0.47\textwidth]{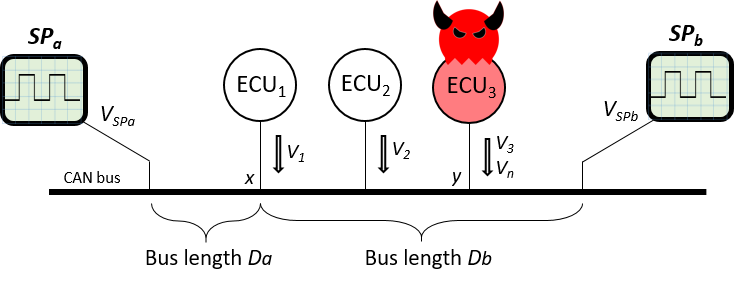}
\caption{Two-Point Fingerprinting Principle}
\label{fig:2pt_concept_highlevel}
\end{center}
\end{figure}

Any given signal being transmitted over a channel will have its propagation affected by the properties of that channel. We utilize these inherent effects as a principle to fingerprint a transmitting ECU. To observe these effects, we consider two points of analog voltage measurements on the bus, sampling points $a$ and $b$, referred to as $SP_a$ and $SP_b$ hereafter, as shown in the Figure~\ref{fig:2pt_concept_highlevel}.

Let ECU\textsubscript{1} be a transmitting node connected to the bus. A physical distance $D_{a}$ separates ECU\textsubscript{1} from and $SP_a$. By the same token, $D_{b}$ is the bus distance to $SP_b$. A message from ECU\textsubscript{1}, as observed from $SP_a$ and $SP_b$, will be affected by the length of the bus it had to travel to reach these sampling points. Therefore, the message observed at $SP_a$ and $SP_b$ will be affected by bus lengths $D_{a}$ and $D_{b}$ respectively. However, a message being transmitted from a different node will be subjected to different lengths of the bus to reach the sample points. By observing the message at two different vantage points, it is possible to extract attributes caused by the channel's influence over the signal, which are correlated to the location of the transmitting node.

The same concept applies to the physical implementation of the CAN bus, which is typically implemented as a pair of twisted wires and therefore it has physical properties associated to it. Furthermore, in practice the physical channel of the CAN bus is implemented as a wiring harness with multiple branches to be able to reach all ECUs of the vehicle. Thus, each ECU\textsubscript{i} has a particular set of distances $\{D_{a_i}, D_{b_i}\}$ to the sampling points $SP_a$ and $SP_b$.

Even though most transmission lines would involve Resistive, Inductive and Capacitive (RLC) effects to any signal being transmitted over it, we focus only on the resistive effects on a simplified representation of the CAN bus to exemplify the concept of using two points of measurements to observe the effects of a bus on the said signal.

\begin{figure}[!htb]
\begin{center}
\includegraphics[width=0.48\textwidth]{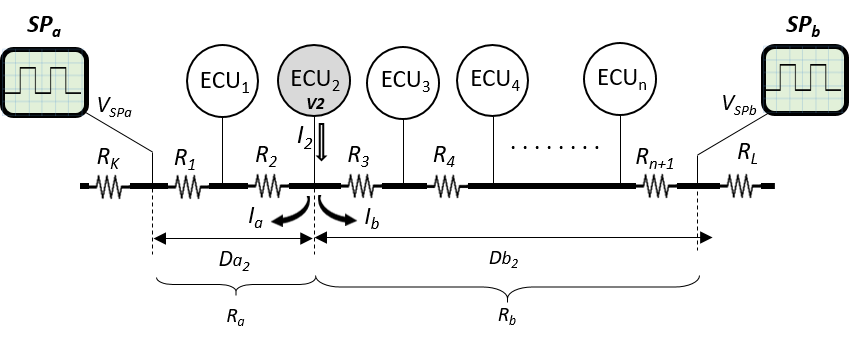}
\caption{Resistive Effects in a CAN Bus}
\label{fig:2pt_concept_detail}
\end{center}
\end{figure}

A more detailed diagram of a channel of the CAN bus is depicted in  Figure~\ref{fig:2pt_concept_detail}, where the wire resistance between the sampling points is denoted by $R_1$ to $R_{n+1}$. $SP_a$ and $SP_b$ are connected close to the extreme ends of the bus with the resistance of the remaining length of the wire denoted as $R_{K}$ and $R_{L}$ respectively. Let ECU\textsubscript{2} be the transmitting node with voltage $V_2$ and current $I_2$ being delivered to the bus. $I_2$ splits into $I_a$ reaching $SP_a$ and $I_b$ reaching $SP_b$ based on the relative resistances of the two branches. From Ohm's law, we can derive the voltage observed at the two sample points as:
\begin{equation}
\label{eq:VSPa1}
V_{SP_a} = I_a \cdot R_K,
\end{equation}
\begin{equation}
\label{eq:VSPb1}
V_{SP_b} = I_b \cdot R_L.
\end{equation}

Since current $I_a$ is the same for $R_1$, $R_2$, and $R_K$, we can describe $V_{SP_a}$ as:
\begin{equation}
\label{eq:VSPa2}
V_{SP_a} = \frac{V_2}{(R_1 + R_2 + R_K)} \cdot R_K.
\end{equation}

Similarly, $V_{SP_b}$ can be determined as: 
\begin{equation}
\label{eq:VSPb2}
V_{SP_b} = \frac{V_2}{(R_3 + R_4 + \cdots + R_{n+1} + R_L)} \cdot R_L.
\end{equation}

From another perspective, it can be observed that the segment of the bus that presents the resistance $R_1 + R_2$ reflects the length of the wire $D_{a_2}$ from the ECU\textsubscript{2} to $SP_a$, and is given by $R_a$. Similarly, the segment $D_{b_2}$ presents the resistance $R_3 + R_4 + ... + R_{n+1}$, which is the length of the wire from ECU\textsubscript{2} to $SP_b$, and is denoted as $R_b$. 

It is assumed that environmental changes affect the resistance of the wire uniformly across its length. This implies that the change in resistance of segment $D_{a_2}$ would reflect a proportional change in the resistance of segment $D_{b_2}$. This change would be similarly exhibited in the voltages $V_{SP_a}$ and $V_{SP_b}$, as voltage is proportional to resistance.

Hence the ratio $\Gamma_{V_2}$ of these observed voltages on the two sample points:
\begin{equation}
\label{eq:ratio1}
\Gamma_{V_2} = \frac{V_{SP_a}}{V_{SP_b}} = \frac{R_K}{R_L} \cdot \frac{(R_{b_2} + R_L)}{(R_{a_2} + R_K)}.
\end{equation}

Equation~\ref{eq:ratio1} can be further generalized to extend to all the ECUs:

\begin{equation}
\label{eq:ratiogeneral}
\Gamma_{V_i} = R_{SP} \cdot \frac{(R_{b_i} + R_L)}{(R_{a_i} + R_K)}\text{,\ where\ } R_{SP} = \frac{R_K}{R_L}.
\end{equation}

Specifically, Equation~\ref{eq:ratiogeneral} shows that for a given transmitter $i$, its ratio $\Gamma_{V_i}$ is independent of the absolute voltage $V_i$ delivered to the bus and is solely dependent on the ratios of the lengths between the ECU and the sampling points. In addition, the constant $R_{SP}$ remains the same for all the ECUs. Moreover, the sum of $R_{a_i}$ and $R_{b_i}$ is a constant as it is the resistance presented by the total length of the wire. Given that no two ECUs are connected to the exact same physical position on the bus, their respective distances $D_{a_i}$ and $D_{b_i}$ are unique, thus making the ratio $\Gamma_{V_i}$ a distinctive fingerprint.

\subsection{Security of Two-Point Fingerprinting}

\begin{figure}[!b]
\begin{center}
\includegraphics[width=0.49\textwidth]{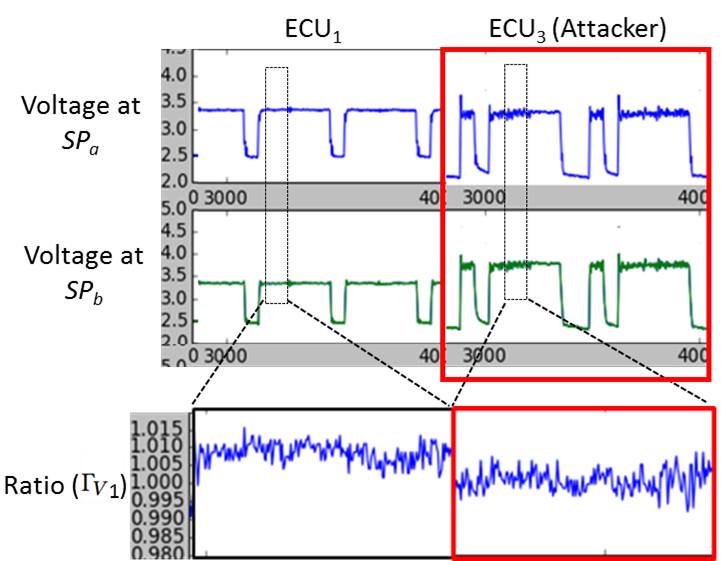}
\caption{Attacker unable to Reproduce an Authentic Ratio $\Gamma_{V_1}$}
\label{fig:voltage_spoofing}
\end{center}
\end{figure}

The voltage ratio computed in the Two-Point fingerprinting scheme identifies the relative location of the ECUs in the bus. The mapping between the bus location and the set of MIDs owned by a given ECU allows for this method to effectively detect both MID and MID-Voltage masquerading.

Consider the case of MID masquerading, as in Figure~\ref{fig:2pt_concept_highlevel}, with ECU\textsubscript{3} as the malicious actor which is connected at location $y$ and attempting to masquerade as ECU\textsubscript{1} which is connected at location $x$. During the attack, ECU\textsubscript{3} would send out MID\textsubscript{1} belonging to ECU\textsubscript{1}, but would output its own voltage $V_3$. For ECU\textsubscript{3} to be effective, it has to match the ratio of the voltages observed at the $SP_a$ and $SP_b$ as the expected ones for ECU\textsubscript{1}. However, in the previous section, we have shown that the ratio $\Gamma_{V_3}$ obtained from the two points are independent of the absolute voltage applied to the bus and are solely dependent on the ratio of the lengths of the bus between the node and the sample points. In other words, MID\textsubscript{1} would not match $\Gamma_{V_3}$, and the MID masquerading attack would be detected.

In the case of MID-Voltage masquerading, ECU\textsubscript{3} would output MID\textsubscript{1} and would try to match ECU\textsubscript{1}'s expected voltages at $SP_a$ and $SP_b$. ECU\textsubscript{3} could output a given $V_{1_a}$ to match ECU\textsubscript{1}'s voltage at $SP_a$, however, it would not satisfy $SP_b$. Similarly, by trying to satisfy $SP_b$'s expectations, ECU\textsubscript{3} would not match $SP_a$'s expected voltage. This is illustrated in Figure~\ref{fig:voltage_spoofing}. In other words, the only for an attacker to craft a voltage signature that results in $\Gamma_{V_1}$ is by satisfying the ratio of voltages observed at $SP_a$ and $SP_b$ simultaneously. This is not possible in an actual car as the attacker has to be in the exact physical location as the authentic transmitting node (as described in the previous section). As a consequence, the MID-Voltage masquerading attack would be detected by the Two-Point fingerprinting. The same concept can be generalized to any ECU\textsubscript{i} in the system. As previously mentioned, the sample points are set up in such a way that there are no ECUs that are equidistant from the two points of measurements.


\subsection{Two-Point Fingerprinting Methodology}

The Two-Point fingerprinting approach proposed in this paper is capable of identifying an ECU from each message that is observed on the CAN bus and is performed in three main phases: In Phase 1, analog samples are collected from the sampling points. Phase 2 is responsible for extracting features that are unique to each ECU. Next, these features are used in Phase 3 to train the classifier and later identifying the transmitting ECU. 

\subsubsection{Phase 1: Analog and Digital Sampling}

\begin{figure}[!htb]
\begin{center}
\includegraphics[width=0.45\textwidth]{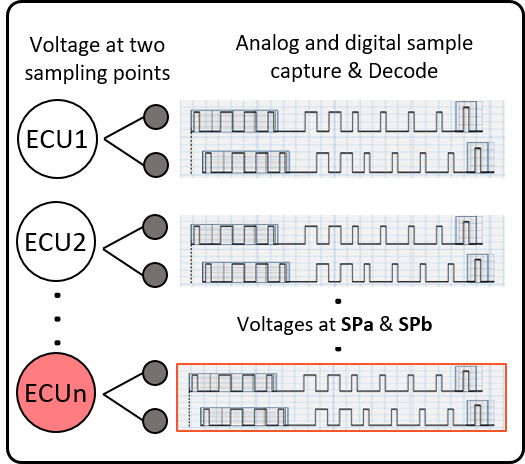}
\caption{Phase 1: Analog and Digital Sampling}
\label{fig:phase1}
\end{center}
\end{figure}

As depicted in Figure~\ref{fig:phase1}, this phase starts with the analog sampling, where sample points $SP_a$ monitors the bus for a message transmission. Whenever a new frame is transmitted, the analog sample acquisition is triggered simultaneously on both $SP_a$ and $SP_b$. This ensures the samples from signals $S(a)$ and $S(b)$ collected from $SP_a$ and $SP_b$ respectively belong to the same message. As discussed in Section~\ref{sec:background}, the MID field of the CAN frame is also used to resolve arbitration, meaning that there might be more than one transmitter active during this field and consequently these samples are ignored. Similarly the samples that belong to ACK field are also ignored. These samples are further used in Phase 2 to extract unique features to fingerprint the transmitting ECU.

The quality of these features is directly dependent on two factors: 1) sampling frequency and 2) sampling resolution. The sampling frequency dictates how many samples are collected during the transmission of a CAN frame and the sampling resolution determines the precision of each measurement.

In addition to the analog samples, a digital capture is performed to decode the MID from the frame being captured. In a typical implementation of CAN bus each ECU is assigned an exclusive set of MIDs. Since we are concerned only with fingerprinting the transmitting ECU, the MID is mapped to the ECU number. For any given captured samples from an ECU\textsubscript{i} a signal vector $\{ECU_i, S(a), S(b)\}$ is created.

\subsubsection{Phase 2: Feature Extraction}

\begin{figure}[!htb]
\begin{center}
\includegraphics[width=0.45\textwidth]{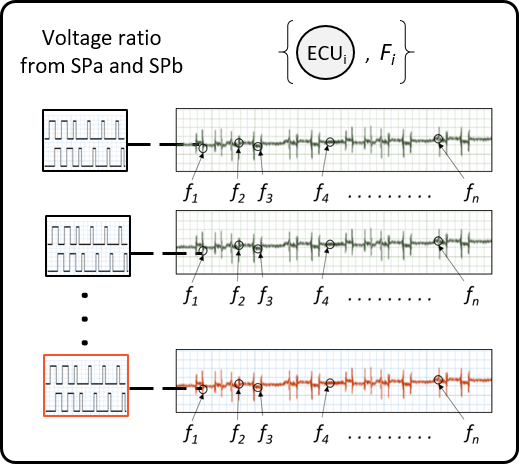}
\caption{Phase 2: Feature Extraction}
\label{fig:phase2}
\end{center}
\end{figure}

The signal vectors $S(a)$ and $S(b)$ capture the effects of the bus on the transmitted signal. As shown in Figure~\ref{fig:phase2}, to derive these effects, a ratio vector $\Gamma_{V}$ is obtained by taking the ratio of $S(a)$ and $S(b)$ as conceptualized in Section~\ref{sec:2p_vf}. Several features become apparent due to the propagation through the physical channel and a number of them have been explored previously in~\cite{CJJPL18, AHTM17}. Some of the features explored are mean, median, standard deviation, root-mean-square (RMS), peak and peak-to-peak values.

Along with the aforementioned features, we can leverage some inimitable features that are unique to the Two-Point fingerprinting scheme. This is only possible because the voltage signal is observed differentially from the two sampling points on the same channel. Instead of looking at the absolute voltage characteristics, the ratio vector captures the effects of the physical channel on the signal. In other words, an observed signal travels different lengths of the bus, and therefore will reach the two sampling points with different voltage characteristics. 

The features discussed in the Section~\ref{sec:comp_related_work} for single-point voltage fingerprinting schemes use the absolute voltage characteristics to fingerprint the transmitting ECU. While it is feasible for Single-Point schemes to fingerprint with high accuracy, it is not possible defend against attackers that are capable of influencing changes to the incident voltage from the transmitter to the physical channel as discussed in Section~\ref{sec:ttm}. All the features extracted in this phase for the Two-Point scheme are explicit products of the relative physical location of the transmitted signal from the sampling points. ECUs are tucked away at various anchored physical locations in a car at the factory. As a consequence, each ECU connects to the bus in a unique location. 

\subsubsection{Phase 3: Fingerprint Training and Identification}

\begin{figure}[htb!]
\begin{center}
\includegraphics[width=0.45\textwidth]{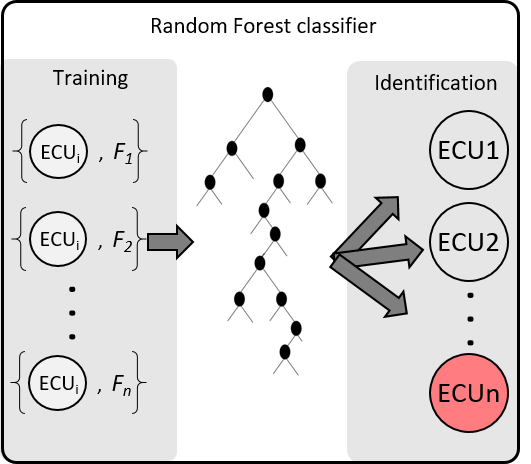}
\caption{Phase 3: Fingerprint Training and Identification}
\label{fig:phase3}
\end{center}
\end{figure}

This phase consists of two parts as shown in the Figure~\ref{fig:phase3}: \mbox{1) Training} and 2) Identification. For every message observed on the bus the features discussed in the previous section are obtained in conjuncture to the ECU label from the captured message frame. During the training phase these features and the ECU labels are used to learn a feature template for each ECU. The supervised learning approach employed here is discussed in detail below. After training, each message is examined and classified against the set of ECUs observed during training. Prior work demonstrates that supervised learning is suitable for fingerprinting ECU with high recall/precision~\cite{CJJPL18}.

\paragraph{Training}
The features extracted in Phase 2 are provided to a supervised learning algorithm with multiple class labels corresponding to each ECU. The ECU labels are used to train the classifier to identify an ECU given a set of features.

Several supervised classifiers were evaluated, and we discovered that Random Forest~\cite{Breiman2001} works best and provides high recall/precision. Notice, however, that other supervised machine learning algorithms can also be employed, and that our proposed scheme does not rely on a particular classification methodology. Prior work in fingerprinting use machine learning classifiers such as Support Vector Machine or Bagged Decision Trees~\cite{CJJPL18}. The key distinction being made in this work is the use of features derived from incorporating two sampling points on the CAN bus, which ultimately allows for better attack coverage (e.g. both MID and MID-Voltage masquerading) than prior work.

As stated in our trust model, this work assumes that the training phase does not contain an active adversary or a misbehaving ECU. Moreover, the ECU labels from Phase 2 are assumed to be correct. Assuming that the attacker does not interfere during training phase is a common assumption~\cite{CJJPL18} when using supervised learning in security applications. In practice, the training can take place before the vehicle is on the road, say, during the manufacturing of the vehicle. While there are other security challenges relating to online training, we consider this out-of-scope of this work.  

\paragraph{Identification} 

The goal of the identification step in Phase 3 is to verify the origin of the received CAN message. To detect if an ECU is masquerading, the ECU label that is produced by the classifier on the received message is compared to MID found in the CAN message. Since all the MIDs that each ECU sends are known ahead of time (MIDs are static and unique to a CAN bus by design), a mismatch between the ECU class label and the observed message indicates the presence of a masquerade attack. The implicit assumption is that all ECU messages observed during identification are also observed during training.

\section{Experimental Evaluation}
\label{sec:exp_results}

The evaluation of the Two-Point fingerprinting method consists of two experimental environments. Our first step is to demonstrate the proof of concept within an ECU testbed, which consist of several microcontrollers that communicate via CAN. The ECU testbed serves as a proving ground for the methodology before validation on a moving vehicle. The Two-Point fingerprinting method was also installed on an actual vehicle, where we successfully demonstrated the feasibility of our novel fingerprint scheme. We confirmed that the performance on the vehicle is similar to the ECU testbed. Due to anonymity reasons, we do not provide the details on the vehicle's make or model.

To validate the concept, the prototype was set up to observe the CAN bus at two points. The analog voltage sample acquisition was performed by using an off-the-shelf USB-based oscilloscope (Digilent Analog Discovery 2). This sampling platform was configured to perform 40Msps (mega samples per seconds) at a 14-bit resolution. The C/C++ APIs provided by the manufacturer was used to trigger and control the USB oscilloscope. More precisely, they were used to provide the CAN protocol decoding and raw sample captures. 

\subsection{ECU Testbed}
\label{subsec:ECU_Testbed}

\begin{figure}[!htb]
\begin{center}
\includegraphics[width=0.49\textwidth]{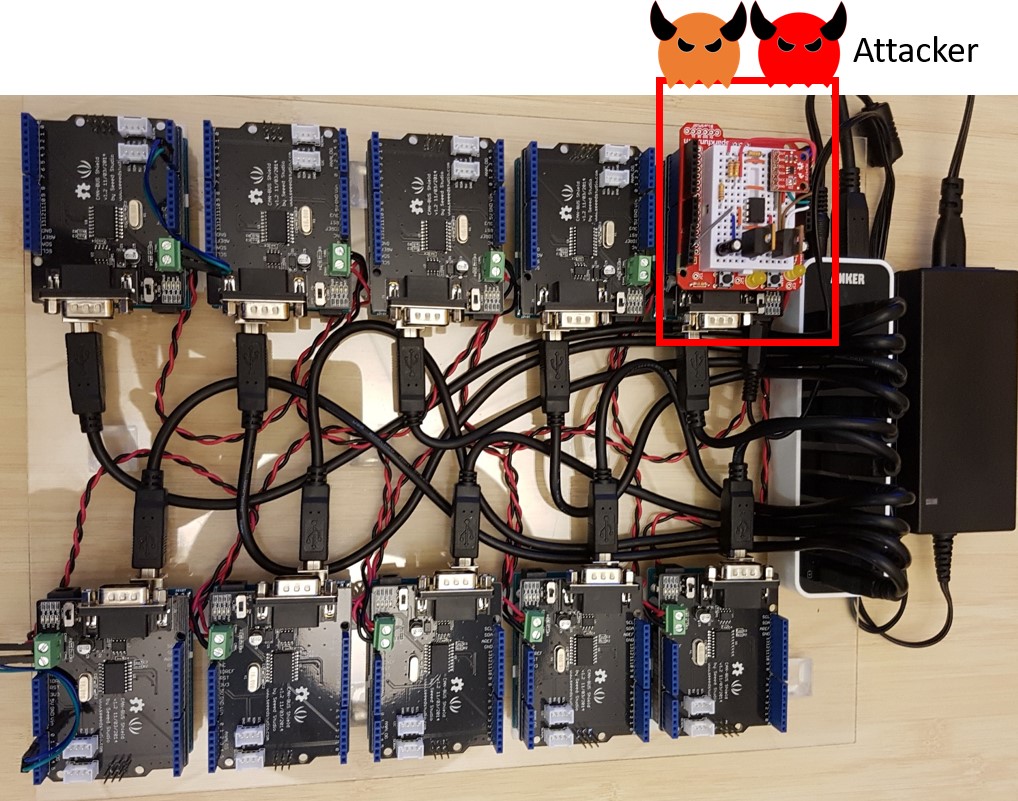}
\caption{ECU Testbed}
\label{fig:ecu_testbed}
\end{center}
\end{figure}

The testbed consists of ten Arduino Uno\texttrademark\  boards with CAN bus shields acting as individual ECUs connected to generate traffic on a single CAN bus, as shown in the Figure~\ref{fig:ecu_testbed}. These ECUs were programmed to periodically transmit their ECU number as the message ID respectively with periodicities ranging from 10 to 40ms. The specific periodicities are based on frequencies observed in real vehicles. 

The fifth ECU on the testbed was repurposed to be the attacker, which is capable of masquerading as any of the other ECUs on the bus. This attacker ECU is also capable of changing the voltage delivered to the bus as discussed in the Section~\ref{sec:attack_module}. The attacker injects messages at a similar periodicities of other ECUs on the CAN network at 10, 20, and 40ms. 

To evaluate feasibility of the Two-Point fingerprinting method on a real car, we performed voltage measurements at two points of the bus. We experimented with multiple CAN networks available in the vehicle. The number of ECUs that are present in a network differ greatly with the make and model of the vehicle. Each CAN network comprises several ECUs, and the number varies depending on the make and model~\cite{MV14}.

\vspace*{-3mm}
\subsection{Training and Testing Configuration}
\label{subsec:training_and_testing_configuration}

Our experimental evaluation consists of a total of about 50-60k message samples (feature sets generated at the end of Phase 2) for both the ECU testbed and for each CAN network on the vehicle.

\subsubsection{Supervised Learning Configuration}

For our experimental evaluation, we used Scikit-Learn's random forest classifier \cite{scikit-learn} with default parameter settings. Prior work \cite{CS17} demonstrates high accuracy with random forest classifier in fingerprinting ECUs at the physical layer. We also considered other classifiers such as Scikit-Learn's SVM and Na\"{i}ve Bayes but concluded that random forest works best.

For every acquired message, forty features are extracted from the ratio vector $\Gamma_{Vn}$. These features are based on statistics of the signal, as used and presented in detail in \cite{AHTM17,CJJPL18}, as well as widths and distances of various peaks and valleys. 

\subsubsection{Training/Testing Split}

The amount of training data necessary for accurate fingerprinting varies, depending heavily on the length of the CAN bus and the distance between ECUs and the sample points. Through a series of empirical evaluations we determined the minimum amount of training data to yield consistent results across multiple runs for both the ECU testbed and in-vehicle. For the ECU testbed evaluation, the training set sizes range from 200 to 20k samples. We observed that 3,598 ($\sim$6\% of total data collected) samples for training is sufficient. Similarly, the for the in-vehicle evaluation, we explored several different training set sizes, ranging from 8k to 20k. The amount of data necessary to produce consistent results is about 12k ($\sim$24\% of total samples), which is about 3.33x more data necessary compared to the ECU testbed evaluation. 

The vehicle feature data collection was conducted over several days to verify the reproducibility between runs. We collected data on the vehicle for several CAN buses and evaluated on a single CAN network at a time. While our experimental evaluation only considers coverage for a single networks, multiple instances of the Two-Point fingerprinting method can be deployed to cover the entire set of CAN networks for a vehicle.

\subsection{Evaluation Metrics}

The classification evaluation metrics for each ECU is based on the evaluation of true and predicted labels, therefore resulting in individual detection rates of true-positives (TP), true negatives (TN), false-positives (FP) and false-negatives (FN). The precision and recall of each ECU are computed respectively as \mbox{$P=TP/(TP+FP)$} and \mbox{$R=TP/(TP+FN)$}.

Precision and recall are consolidated as F1-Score, as follows:
\begin{equation}
\label{eq:f1score}
\text{F1-Score} = 2 \cdot \frac{P \cdot R}{P + R}.
\end{equation}

The results reported in the next section refers to the average F1-score of all ECUs present in the experimental setting. Worth mentioning that the number of ECUs in the testbed was not the same as in the real vehicle.

\subsection{Results}
\label{subsec:results}
\begin{figure}[!htb]
\begin{center}
\includegraphics[width=0.49\textwidth]{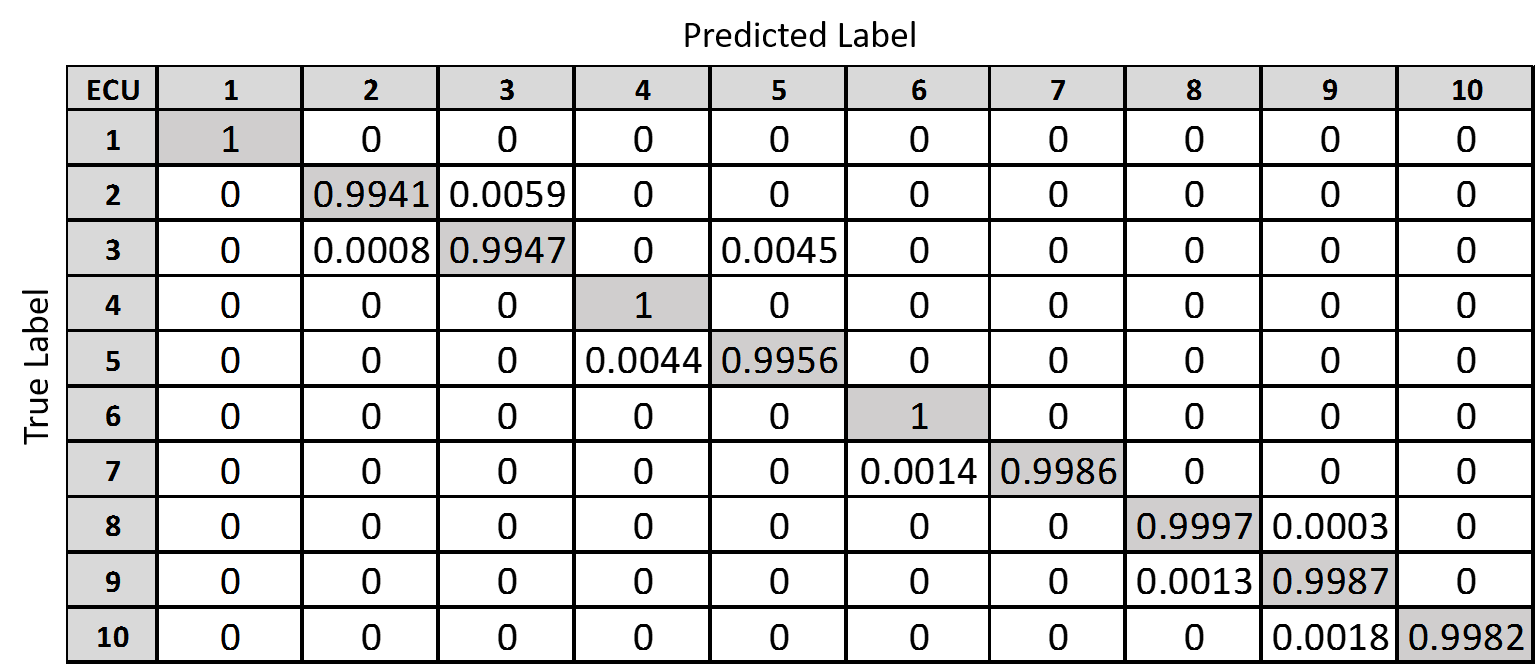}
\caption{Confusion Matrix, with 6\% Training (3,598 samples) and 94\% Testing (56,384 samples).} 
\label{fig:conf_mat_ccs_6p}
\end{center}
\end{figure}

Our experimental evaluation consists of two attack strategies: a) MID Masquerading and b) MID-Voltage Masquerading. The first case evaluates our proposed method under attackers who injects messages without modifying voltage and demonstrate an F1-Score above 0.994. The latter case demonstrates that under a stronger attacker assumption, where the attacker masquerades both MID and voltage levels, the Two-point Voltage Fingerprinting method is capable of detecting the masquerade attack with the same level of accuracy.
\paragraph{MID Masquerading Attacks} The Two-Point fingerprinting was evaluated using the ECU testbed. A confusion matrix summarizing the results is shown in Figure~\ref{fig:conf_mat_ccs_6p}. Notice that the average F1-score for individual ECU identification is 0.9980. The worst F1-score in the case of ECU\textsubscript{3} was 0.9943, which is still considerably high. For a dataset of 59,982 samples, only 6\% of them (3,598) were used for training. Even this small percentage of training data was sufficient to produce the aforementioned F1-scores. This indicates that a training with a large number of samples is not required to achieve high accuracy.

\begin{figure}[!htb]
\begin{center}
\includegraphics[width=0.49\textwidth]{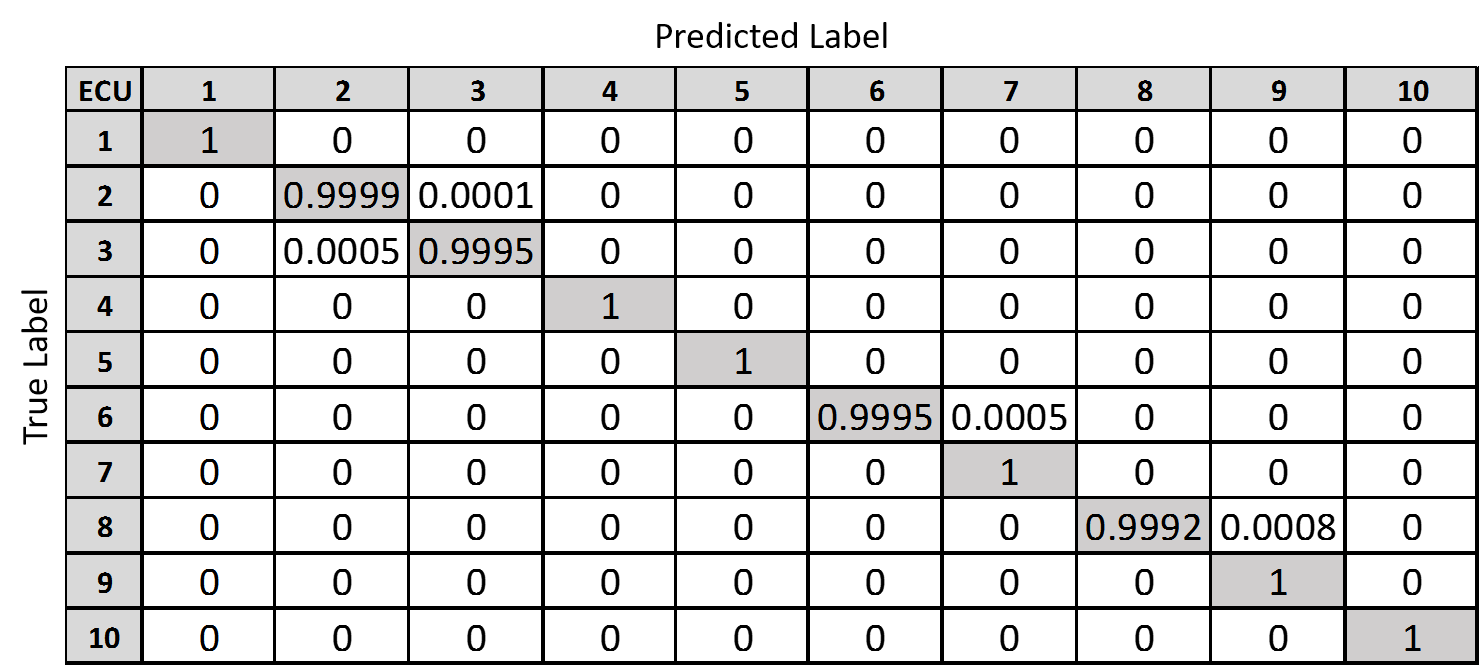}
\caption{Confusion Matrix, Stratified 10-Fold Cross Validation, n=59,982.} 
\label{fig:conf_mat_ccs_10f}
\end{center}
\end{figure}

We also evaluated it under a stratified 10-fold cross validation as shown in Figure~\ref{fig:conf_mat_ccs_10f}, where 90\% of the data is used for training and 10\% is used for testing under ten non-overlapping folds. The average F1-score for all ECU class labels is 0.9998 with the worst performing ECU label at 0.9996. The results above indicate that the proposed method is able to fingerprint ECUs at a high level of accuracy.

In addition, we performed similar experiments on CAN networks of an actual vehicle. The results demonstrated that the Two-Point fingerprinting scheme is able to achieve high F1-scores in a real vehicular environment, for both parked and driving states. During those tests, the best and worst F1-scores were 0.9936 and 0.9741, respectively.

\paragraph{MID-Voltage Masquerading Attacks}

In order to evaluate the accuracy of detecting MID-Voltage masquerading attacks, we used ECU\textsubscript{5} as the attacker node in our ECU testbed. ECU\textsubscript{5} launched about 600 attack messages against each of the randomly selected ECUs (ECU\textsubscript{3}, ECU\textsubscript{7}, and ECU\textsubscript{8}). Besides MIDs, the attacks consisted of injecting voltages to match those of the target ECUs. Specifically, CANH levels of 3.2185V, 3.3114V, and 3.405V, for ECU\textsubscript{3}, ECU\textsubscript{7}, and ECU\textsubscript{8}, respectively. 

\begin{figure}[!htb]
\begin{center}
\includegraphics[width=0.48\textwidth]{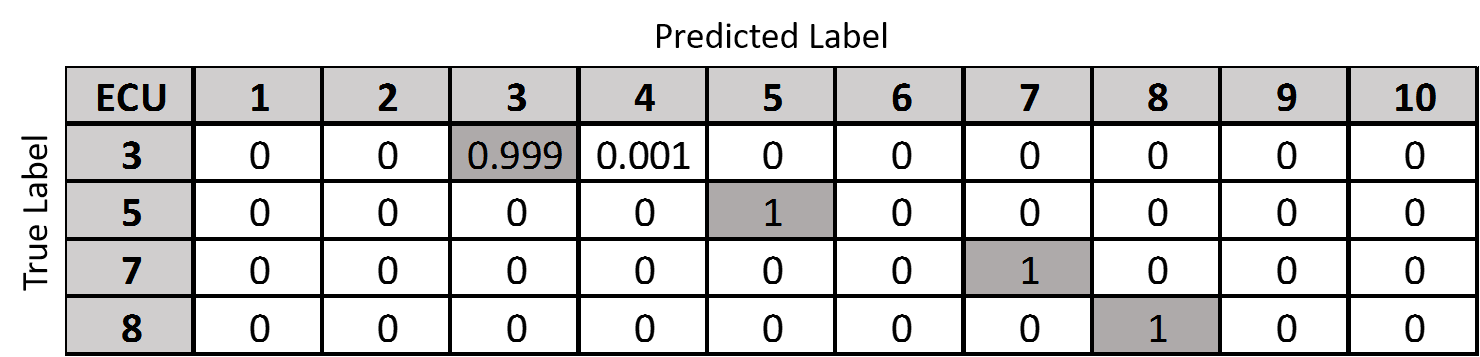}
\caption{Confusion Matrix for MID-Voltage Masquerade Attacks} 
\label{fig:attack_conf_mat}
\end{center}
\end{figure}

The confusion matrix listed in Figure~\ref{fig:attack_conf_mat} was created using the ECU labels returned by the Two-Point fingerprinting scheme. It can be seen that even though ECU\textsubscript{5} was trying to masquerade as the other ECUs, it was still correctly identified as ECU\textsubscript{5} with near perfect accuracy. In addition, the identification accuracy of the targeted ECUs remained high, and therefore unaffected.

\section{Related Work and Discussion}
\label{sec:comp_related_work}

MID masquerading attacks are currently the most prevalent and pervasive cyberattacks in the automotive industry. These are also the most studied attacks in literature. The Two-Point fingerprinting method is a non-intrusive method. In other words, it does not require modification to any of the ECU hardware or software/firmware. It can be deployed at both manufacturing time and retrofitted into existing vehicles. The implementation during the manufacturing of the vehicle has the benefit of allowing for the identification of optimal tap points for the voltage measurements. After market installation is also possible, granted that it is possible to tap into two different locations of the CAN bus. In both cases, the ideal locations for the sampling points are at the extremities of the CAN bus, so that we can observe maximum influence of the channel on the electrical signals.

Real world embodiment of vehicular CAN buses are not always linear. Rather, they are more akin to a tree.  Even considering this practical limitation, we have demonstrated the high accuracy of the proposed method. The experiments performed on a real car used OBD-II port as the first measuring point, whereas the second sampling point was connected to an ECU in the trunk of the car. In the testbed, the bus is linear with the sampling points at its two extremeties. This practical aspect is the reason behind the slight variation in the F1-score observed for the worst case between the testbed and the car. Additionally the number of ECUs in the car is not the same as in our ECU testbed\footnote{The exact number of ECUs in the test vehicle cannot be revealed due to confidentiality.}.

One of the initial proposals to detect such attacks/intrusions in the automotive environment is reported in~\cite{HKD11}. It describes a scheme to detect MID misuse, which would only be effective if all nodes are connected directly to a gateway. Some high level ideas on utilizing physical level characteristics are discussed as well. This work also describes the concepts of using message frequency analysis to detect MID masquerading attacks. However, it does not provide means to cover messages that are sent sporadically rather than periodically. Similar frequency based attack detection mechanisms~\cite{CS16} rely on mapping the ECU based on the arrival times of their transmitted messages. While these schemes can detect MID masquerading attacks in periodic messages, they are unsuitable for sporadic messages. In the Two-Point fingerprinting, detection is agnostic to the messaging frequency. As long as it has been trained across the ECUs, the periodicity of their messages does not play a role in detection.

Information theory models have been proposed in~\cite{MA11} to monitor disturbances on the expected message frequencies and correlated events. On one hand, abrupt changes in the expected entropy can be flagged as an attack, but that also occur during normal operation of the vehicle. On the other hand, attacks causing minor entropy impact would go undetected. In summary, methods based on pattern/entropy analysis have an expected effectiveness limitation due to higher false positives and negatives. Other methods~\cite{SKK16,MBCSP17} proposed to utilize message interval times to detect message injection, but they still fall short in covering sporadic messages. Machine learning based approaches~\cite{KK16} can detect simple data spoofing attacks by examining the message contents, but require attack samples in its training phase. The Two-Point fingerprinting does not consider the contents of the messages as a distinguishing feature, thus avoiding false positive detections due to variations in the message payload.



Several Intrusion Detection Systems (IDS) have been proposed based on the analysis of the physical channel. In~\cite{CS16}, a clock-skew based IDS is utilized to identify MID masquerading attacks by fingerprinting the timing characteristics of each ECU. ECU fingerprinting has also been proposed in~\cite{MG14,CS17,AHTM17,CJJPL18}, which uses the voltage profile of each ECU as the ground truth. Though each of them propose a different variant to learn and classify observed patterns, they are susceptible to MID-Voltage masquerading attacks, as discussed in Section~\ref{sec:attack_module}.

In comparison with related work, single-point fingerprinting methods such as VoltageIDS~\cite{CJJPL18}, Viden~\cite{WHSJJD16,CS17} use voltage signals characteristics to identify masqueraded messages. Those approaches assume that the attacker can exclusively masquerade MIDs, but not ECU voltages. Our work demonstrates that an attacker can mimic voltage profiles of other ECUs with low-cost commodity hardware. While some of the features and identification method are similar to the work presented here, our approach is resilient against a more capable attacker that can masquerade both MIDs and voltages. Also the differential nature of sampling the CAN bus described in this paper makes the Two-Point fingerprinting scheme more robust to common-mode operational noise.

\section{Conclusions}
\label{sec:conclusions}

Attackers persistently craft novel attacks to bypass existing security mechanisms. Automotive systems are no exception. There are attacks that can jeopardize vehicle safety. The lack of authentication in the "de-facto" CAN bus has been the underlying issue that has been exploited. As a countermeasure, researchers have proposed ECU voltage fingerprinting schemes to detect attacks and therefore address this limitation of CAN.

Voltage impersonation attacks have been reported in the literature as a bypass of the aforementioned schemes. In this paper, we demonstrate the feasibility of performing \mbox{MID-Voltage} masquerading attacks using the CAN message ID along with victim ECU's voltage, therefore circumventing single-point fingerprinting schemes.

In the automotive domain, a hardware Trojan that is capable of spoofing voltage profiles of authentic ECUs would remain undetected by Single-Point fingerprinting methods. While no computing system can be completely secure, our proposed Two-Point voltage fingerprinting approach is capable of detecting both MID and \mbox{MID-Voltage} masquerading attacks with an overall \mbox{F1-score} of above 99.4\% for both ECU testbed and actual vehicles.

This new approach raises the bar over existing fingerprinting mechanisms, therefore increasing the effectiveness of intrusion detection systems. This serves as a solid foundation for improved cybersecurity, and consequently, increased safety in autonomous vehicles.


\bibliographystyle{IEEEtran}
\bibliography{refs}

\end{document}